\documentclass[12pt]{article}
\usepackage{amssymb,amsmath}
\usepackage[hypertex]{hyperref}

\usepackage{latexsym}

\numberwithin{equation}{section}

\def\coeff#1#2{\relax{\textstyle {#1 \over #2}}\displaystyle}

\def\IR{\mathbb{R}}

\def\Neql#1{{\cal N}\!=\!{#1}}

\def\cB{{\cal B}}

\def\cF{{\cal F}}
\def\cG{{\cal G}}
\def\cM{{\cal M}}
\def\cN{{\cal N}}

\newcommand{\be}{\begin{equation}}
\newcommand{\ee}{\end{equation}}
\newcommand{\bea}{\begin{eqnarray}}
\newcommand{\eea}{\end{eqnarray}}

\begin{document}

\begin{titlepage}

\begin{flushright}
\end{flushright}

\bigskip
\bigskip
\centerline{\Large \bf Supersymmetric Solutions in Six Dimensions:}
\smallskip
\centerline{\Large \bf A Linear Structure}
\medskip
 \bigskip
\centerline{{\bf Iosif Bena$^1$, Stefano Giusto$^{2,3}$, Masaki Shigemori$^4$ and Nicholas P. Warner$^5$}}
\bigskip
\centerline{$^1$ Institut de Physique Th\'eorique, }
\centerline{CEA Saclay, CNRS-URA 2306, 91191 Gif sur Yvette, France}
\bigskip
\centerline{$^2$ Dipartimento di Fisica ``Galileo Galilei,''}
\centerline{ Universit\`a di Padova, Via Marzolo 8, 35131 Padova, Italy}
\bigskip
\centerline{$^3$ INFN, Sezione di Padova, Via Marzolo 8, 35131 Padova, Italy}
\bigskip
\centerline{$^4$Kobayashi-Maskawa Institute for the Origin of Particles and the Universe,}
\centerline{Nagoya University, Nagoya 464-8602, Japan}
\bigskip
\centerline{$^5$ Department of Physics and Astronomy}
\centerline{University of Southern California} \centerline{Los
Angeles, CA 90089, USA}
\bigskip
\centerline{{\rm iosif.bena at cea.fr,~stefano.giusto at pd.infn.it,} }
\centerline{{\rm shige at kmi.nagoya-u.ac.jp,~warner at usc.edu } }
\bigskip
 
\bigskip \bigskip

\begin{abstract}
The equations underlying all supersymmetric solutions of six-dimensional
minimal ungauged supergravity coupled to an anti-self-dual tensor
multiplet have been known for quite a while, and their complicated
non-linear form has hindered all attempts to systematically understand
and construct supersymmetric solutions. In this paper we show that, by suitably
re-parameterizing these equations, one can find a structure that allows
one to construct supersymmetric solutions by solving a sequence of
linear equations. We then illustrate this method by constructing a new
class of geometries describing several parallel spirals carrying D1, D5
and P charge and parameterized by four arbitrary functions of one
variable. A similar linear structure is known to exist in five
dimensions, where it underlies the black hole, black ring and
corresponding microstate geometries.  The unexpected generalization of
this to six dimensions will have important applications to the
construction of new, more general such geometries.
 
\end{abstract}

\end{titlepage}


\section{Introduction}
\label{sect:Intro}

Much of the remarkable progress in constructing and classifying
three-charge, supersymmetric solutions in five-dimensions was made
possible by the observation that the underlying equations
\cite{Gauntlett:2002nw,Gutowski:2004yv,Bena:2004de}, contrary to initial
appearances, are \emph{linear} \cite{Bena:2004de}: the whole system can
be reduced to solving the equations of four-dimensional Euclidean
electromagnetism with sources. This enabled the discovery of the
bubbling transition \cite{Bena:2005va,Berglund:2005vb}, which lies at
the core of the construction of supersymmetric black-hole microstate
geometries, and has also led to rich new classes of BPS\footnote{In this
paper we use the terms ``BPS'' and ``supersymmetric'' interchangingly,  as is common
in the literature on supersymmetric supergravity solutions.} solutions.

It is a natural and important question to ask whether this linearity
extends to higher dimensional supergravities and, more particularly, to
the six-dimensional supergravity theory \cite{Gueven:2003uw,
Gutowski:2003rg, Chamseddine:2003yy, Cariglia:2004kk, Jong:2006za,
Akyol:2010iz} whose reduction yields the five-dimensional, $\cN=2$
ungauged supergravity theory where the linear structure is known to
exist. Given that all supersymmetric, asymptotically $AdS_3 \times S^3$
solutions that are dual to the D1-D5 system belong to this class, the
significance of a linear structure in the six-dimensional theory can
hardly be overemphasized. The generic BPS geometry in six dimensions
would, of course, differ from its five-dimensional counterpart in that
it can depend non-trivially upon the compactification direction and this
could easily lead to non-linearities.  On the other hand, if the
equations were still linear it would not only make the construction of
explicit solutions much more straightforward but it would also enable a
much more precise analysis of the moduli space of such solutions.

One of the key aspects of the fuzzball proposal\footnote{See, for example,  \cite{Mathur:2005zp, Bena:2007kg, Mathur:2008nj, Balasubramanian:2008da, Skenderis:2008qn, Chowdhury:2010ct} for reviews.} is that the microstate geometries of a black hole constructed in four-dimensional supergravity cannot be properly described and counted unless one works in a higher dimensional supergravity. Indeed, the singularity theorems of four-dimensional general relativity tell us that there are essentially no four-dimensional microstate geometries.  On the other hand, the structure of microstate geometries becomes far richer in five and six dimensions.  The bubbling transition that resolves the singularity and gives the smooth microstates of the four-dimensional D6-D2-D2-D2 black hole naturally emerges in five-dimensional supergravity  \cite{Bena:2005va,Berglund:2005vb,Saxena:2005uk}. 

It is true that the easiest-to-construct five-dimensional microstate geometries have a tri-holomorphic $U(1)$ invariance, and thus descend to multi-center black-hole solutions in four-dimensions \cite{Denef:2000nb,Bates:2003vx}.   Such microstate geometries may therefore be thought of as being four-dimensional but they are singular from the four-dimensional perspective. Moreover, solutions with this  $U(1)$ invariance have a finite-dimensional moduli space, and hence are ``too rigid'' to account for entropy of the black hole \cite{Bena:2006is,Bena:2007qc,deBoer:2008zn}\footnote{Despite this, these solutions can sample the typical sector of the underlying conformal field theory \cite{Bena:2006kb, Bena:2007qc,deBoer:2008zn,deBoer:2009un} and give us insights on how a typical microstate geometry looks.}. Typical microstates will not  have this $U(1)$ invariance and hence will be intrinsically  five-dimensional, at least.   Indeed, a lot of recent work suggests that, while the essential geometric transition can be made in five dimensions, it is only in six, or more, dimensions that microstate geometries have a rich enough moduli space to sample black-hole microstates with sufficient density to obtain a semi-classical thermodynamic model.

One proposal to obtain such solutions has been to place arbitrarily-shaped supertubes in three-charge bubbling solutions, and thus obtain infinite-dimensional families of three-charge black-hole microstate geometries \cite{Bena:2010gg}. Via the so-called ``entropy enhancement" mechanism,  these supertubes are expected to have more entropy than supertubes in flat space \cite{Bena:2008nh,Bena:2008dw}.  A rough estimate is that the entropy of normal supertube scales like $\sqrt{Q^2}$, while the entropy of a single enhanced supertube scales like $\sqrt{Q^{5/2}}$, which is a significant increase but still below the black hole scaling, $\sqrt{Q^3}$.

All the supertube entropy counts are done essentially using a Cardy-type formula, in which the central charge is the number of arbitrary continuous functions that determine the supergravity solution. Hence, the obvious way to bypass the $\sqrt{Q^{5/2}}$ bound is to find solutions that depend on a very large number of functions of one variable. One possibility is to consider multiple wiggly supertubes, possibly rotating in opposite directions. In the probe approximation, such tubes have infinite entropy, and there is hope that upon back-reaction this entropy will grow as $\sqrt{Q^3}$ \cite{Bena:2008nh}. Unfortunately, the back-reaction calculation is even more complicated than for a single wiggly supertube \cite{Bena:2010gg} (which is not particularly simple either) and the fact that one  knows the Green's functions  for only very specific ambipolar backgrounds makes the  extension of this work to general backgrounds all the more difficult.

However, there is now the possibility of a new, even richer candidate
microstate geometry: The \emph{superstratum} \cite{Bena:2011uw}.  Based
upon a careful analysis of the supersymmetry structure of three-charge
brane configurations, it was argued in \cite{Bena:2011uw} that
configurations with three electric charges can be given additional
dipole charges to create a new class of branes whose shape is given by
functions of \emph{two} variables. These configurations are
intrinsically six-dimensional, and have $16$ supersymmetries locally (in
each of the tangent spaces of the superstratum) but globally preserve
only the four supersymmetries preserved by their three electric charges.

What distinguishes the superstratum from other exotic branes
configurations \cite{deBoer:2010ud} is that in the D1-D5-P duality frame
all the branes that form the superstratum can be given a geometric
description and the superstratum supergravity solution is very likely to
be \emph{smooth}. Thus the superstratum, if it exists in as full
generality as expected, will give rise to microstate geometries that
depend upon functions of two variables and these will necessarily
provide an even richer sampling of microstates of the black hole. In
particular, it is quite plausible that quantizing the fluctuations in
one direction will give rise to a parametrically-large number of
functions of one variable, which in turn will increase the central
charge in the entropy counting, and perhaps give rise to an entropy that
grows like $\sqrt{Q^3}$

As explained in \cite{Bena:2011uw}, a superstratum can become \emph{rigid} when the functions of two variables that determine its shape become functions of only one variable. Such a rigid superstratum is no longer intrinsically six-dimensional  but five dimensional, and has eight supersymmetries.  It is, in fact, a supertube. 

The highly non-trivial aspect of the conjecture in  \cite{Bena:2011uw} is that a superstratum can be given arbitrary shape modes that are functions of two variables  and yet remain smooth. Indeed, the construction of the superstratum was based upon making two successive supertube transitions on the original set of D-branes and the functions of two variables emerge because one can make the two transitions independently so that at each point on the first supertube profile one can  choose an independent profile for the second supertube.  The regularity of the end result comes about for the same reason that the D1-D5 supertube is smooth  \cite{Lunin:2001jy,Lunin:2002iz}.

While this conjecture is very plausible it still remains to be proven and one way to achieve this is to construct it explicitly.  The general superstratum has D1, D5 and P charges, and lives in type IIB supergravity compactified on $T^4$ or $K3$. Its solution is therefore a supersymmetric solution of minimal, ungauged, six-dimensional supergravity coupled to one or several anti-self-dual tensor multiplet. Fortunately, the equations governing all these supersymmetric solutions have been written down in \cite{Cariglia:2004kk}, generalizing earlier work on the minimal supergravity \cite{Gutowski:2003rg}.  Thus we already know what equations the superstratum solution will solve. However, as anybody familiar with \cite{Gutowski:2003rg,Cariglia:2004kk} knows, finding an intrinsically six-dimensional solution to these complicated equations is  highly non-trivial (see for example \cite{Ford:2006yb} for one such solution).

Our purpose in this paper is to take a step towards the construction of
the superstratum solution by simplifying the non-linear equations of
\cite{Gutowski:2003rg,Cariglia:2004kk} and showing that in fact they can
be solved in a linear procedure! The linear structure we uncover here is
similar to the one that governs all supersymmetric solutions of
\emph{five-dimensional} ungauged supergravity
\cite{Bena:2004de}. Indeed, the six-dimensional BPS system also has some
remarkable similarities with the generalized linear system, discovered
in \cite{Bena:2009fi}, describing families of non-BPS solutions with
floating branes.

The fact that the six-dimensional BPS problem is linear means that the
various elements of a complicated superstratum solution will only
interact in certain way, and in particular that multiple superstrata can
be superposed, and will most likely not destroy each other's wiggles (as
might happen in a system governed by non-linear equations), much like
multiple wiggly supertubes constructed in five-dimensional supergravity
do not hinder each other's oscillations because of the linearity of the
five-dimensional solution.  Hence, once a superstratum solution is
obtained we expect to be able to superpose them and to study the moduli
space of multiple superstrata.

We will leave the construction of the general superstratum for future work and begin here by exposing the linear structure of the six-dimensional equations.  Then, to illustrate the power of the linear system and to make a first step toward the superstratum, we construct a solution describing multiple interacting supertube spirals with D1, D5 and P charge.  These are the charges of a superstratum and by giving them an arbitrary profile we are making the first of the two supertube transitions  that lead to the superstratum \cite{Bena:2011uw}. 

Section 2 starts by summarizing the essential results about the six-dimensional supergravity theories  \cite{Gutowski:2003rg,Cariglia:2004kk}, and then goes on to re-write the system of equations in a manner that exposes the underlying linearity.  Section 3 contains some further simplifications to the system of equations and identifies the electric and magnetic fluxes.    Section 4 contains some examples of new solutions that can be generated using this linear structure and Section 5 points to a broad new class of solutions that should be within relatively easy reach for analysis.  Section 6 contains our conclusions.

\section{Supersymmetric backgrounds}
\label{sect:susyconds}

Minimal ungauged $\Neql 1$ supergravity in six dimensions has a bosonic field content  consisting of a graviton and a  two-index tensor gauge field, $B_{\mu \nu}$, whose field strength, $G$, is required to be self-dual.  If one reduces this to five dimensions one obtains $\Neql 2$ supergravity coupled to one vector multiplet.  We want to consider the slightly more general, ``non-minimal''  theory in six dimensions whose dimensional reduction yields  $\Neql 2$ supergravity coupled to two vector multiplets.  The corresponding $\Neql 1$ supergravity in six dimensions is simply the minimal theory coupled to an extra  anti-self-dual  tensor multiplet.  The complete bosonic field content is then a graviton, a \emph{general} two-index tensor gauge field, $B_{\mu \nu}$, and a dilaton, $\phi$.  

The conditions for the most general supersymmetric solutions of the minimal $\Neql 1$ gauged supergravity in six dimensions were first obtained in \cite{Gutowski:2003rg}.  In  \cite{Cariglia:2004kk}, this was then generalized to an even larger class of six-dimensional  $\Neql 1$ supergravity theories coupled to vector multiplets.  One can then extract the supersymmetry conditions 
for the non-minimal ungauged  $\Neql 1$ theory of interest here by setting all the vector fields and gauge couplings to zero in   \cite{Cariglia:2004kk}.  

The first part of the supersymmetry analysis implies that the solution
must have a null Killing vector and that all fields must have vanishing
Lie derivative along that direction.  We will choose a coordinate, $u$,
along this Killing vector and then all fields will be
$u$-independent. However, the fields will, in general, depend upon the
other five coordinates.  The null Killing vector introduces a $2+4$
split in the geometry and so it is natural to introduce a second
retarded time coordinate\footnote{Note that we are reversing the
conventions of \cite{Gutowski:2003rg,Cariglia:2004kk}, by interchanging
the roles of the coordinates $u$ and $v$.}, $v$, and a four-dimensional,
and generically $v$-dependent, spatial base, $\cB$, with coordinates
$x^m$, $m=1,\dots,4$.

\subsection{The metric}

The six-dimensional metric is:
\begin{equation}
ds^2 =  2 H^{-1} (dv+\beta) \big(du + \omega ~+~ \coeff{1}{2}\, \cF\, (dv+\beta)\big) ~-~  H \, ds_4^2\,, \label{sixmet}
\end{equation}
where the metric on the four-dimensional base, $\cB$, is written in
terms of components as:
\begin{equation}
 ds_4^2~=~  h_{mn} dx^m dx^n \,. 
 \end{equation}
In \eqref{sixmet}, $\beta=\beta_m dx^m$ and $\omega=\omega_m dx^m$ are regarded as
1-forms on $\cB$. The functions $H$ and $\cF$, the 1-forms $\beta$ and $\omega$,
and the metric $h_{mn}$ all depend on $v$ and $x^m$ but not on $u$.

The supersymmetry conditions imply that the base  is almost hyper-K\"ahler  in that there are three anti-self-dual $2$-forms, 
\begin{equation}
 J^{(A)}  ~\equiv~ \coeff{1}{2}\,   {J^{(A)}}_{mn}  \, dx^m  \wedge dx^ n \,,
 \end{equation}
that satisfy the algebra: 
\begin{equation}
{J^{(A)}}{}^m{}_p {J^{(B)}}{}^p{}_n = \epsilon^{ABC}\,{J^{(C)}}{}^m{}_n ~-~  \delta^ {AB} \,\delta^m_n \,.
\label{Jalg} 
 \end{equation}
These forms are also required to satisfy the differential identity:
\begin{equation}
\tilde d J^{(A)}  ~=~    \partial_v \big (\beta \wedge J^{(A)})  \label{Jcond} \,,
 \end{equation}
where $\tilde{d}$ is the exterior derivative restricted to $\cB$, which
acts on a $p$-form, $\Phi \in \Lambda^p ({\cal B})$, by:
\begin{eqnarray} 
 \Phi &=& {1 \over p!} \, \Phi_{m_1 \dots m_p} (x,v) \, dx^{m_1} \wedge \ldots \wedge  dx^{m_p} \,, \\
 {\tilde{d}} \Phi &\equiv&  {1 \over (p+1)!} \, (p+1) \, {\partial \over \partial x^{[q}} \Phi_{m_1 \dots m_p]} \,  dx^q \wedge  dx^{m_1}
 \wedge \ldots \wedge dx^{m_p}\,.
\end{eqnarray} 
Also, $\partial_v\Phi$ denotes the Lie derivative of a quantity $\Phi$
with respect to the tangent vector $\partial \over \partial v$.  We will
also use the notation $\dot{\Phi}\equiv \partial_v \Phi$.  Note that we
are using the conventions of \cite{Gutowski:2003rg} in the definition of
the $J^{(A)}$ and that one must make the replacement $J^{(A)} \to -
J^{(A)}$ in order to go to the conventions of \cite{Cariglia:2004kk}.

There is a remark after Equation (4.13) in \cite{Cariglia:2004kk} that outlines a simple computation that purportedly leads to another differential constraint on the $J^{(A)}$.  However one can use the explicit spin connections given in  \cite{Gutowski:2003rg} to perform this computation and we find, exactly as noted in \cite{Gutowski:2003rg}, that this computation simply yields (\ref{Jcond}) and there are no other constraints on $J^{(A)}$.

Define the anti-self-dual $2$-forms, $\psi$ and $\hat \psi$, by:
\begin{equation}
\psi  ~\equiv~ H \, \hat \psi  ~\equiv~  \coeff{1}{16} \, H\,  \epsilon^{ABC}  \, J^{(A)}{}^{mn} \dot {J}^{(B)}{}_{mn} \, J^{(C)} \label{psidefn}  \,,
\end{equation}
Note that  $\hat \psi$ is anti-self-dual and depends solely upon the  almost hyper-K\"ahler structure on the base.  It should be noted that both references \cite{Gutowski:2003rg} and  \cite{Cariglia:2004kk} use the same definition of $\psi$ and so, given our choice in (\ref{Jalg}),  we have the same sign conventions for $\psi$ as in  \cite{Gutowski:2003rg}  and that one should send $\psi \to -\psi$ to go to the conventions of \cite{Cariglia:2004kk}.

Following \cite{Gutowski:2003rg},  we introduce the frames, $\{e^+,e^-,e^a\}$ in which the metric takes the form: 
\begin{equation} 
ds^2 ~=~   2 e^+ e^- ~-~ \delta_{ab}\, e^a  \, e^b \,; 
\end{equation} 
\begin{equation} 
e^+ \equiv H^{-1} \big(dv + \beta \big)  \,,  \qquad
e^- ~\equiv~  du +\omega   + \coeff{1}{2} \,{\cal F} H \, e^+  \,, \qquad 
e^a = H^{1 \over 2} \tilde{e}^a{}_m dx^m \,,  \label{eqn:frames}
\end{equation} 
and we use the orientation 
\begin{equation} 
\epsilon^{+-1234}  ~=~ \epsilon^{1234} ~=~ +1 \,.
\end{equation} 
In $d$ dimensions, we define the Hodge star $*_d$ to act on a $p$-form
as
\begin{align}
 *_d\, (dx^{m_1}\wedge\cdots\wedge dx^{m_p})
 ={1\over (d-p)!}
 dx^{n_1}\wedge\cdots\wedge dx^{n_{d-p}}\,
\epsilon_{n_1\dots n_{d-p}}{}^{m_1\dots m_p}.
\end{align}
It is convenient to introduce the Kaluza-Klein covariant differential
operator, $D$, defined by:
\begin{equation} 
D \Phi ~\equiv~ {\tilde{d}} \Phi ~-~ \beta \wedge {\dot{\Phi}}.  \label{betaFS}
\end{equation} 
The vector field, $\beta$, is then required to satisfy the self-duality condition:
\begin{equation} 
D \beta ~=~  *_4 D \beta \,. \label{betacond}
\end{equation} 
Note that if one acts with $\tilde d$ on both sides of (\ref{Jcond}), and then re-uses the equation, one obtains the integrability condition:  $\partial_v(D\beta \wedge J^{(A)}) =0$, which follows from the self-duality of  $D \beta$ and the anti-self-duality of $J^{(A)}$.

The first step in obtaining a supersymmetric background is to select  an almost hyper-K\"ahler base, whose almost complex structures satisfy (\ref{Jalg}) and  (\ref{Jcond}), and then obtain a vector field, $\beta$, satisfying (\ref{betacond}).  We will show below that this represents the only non-linear aspect of finding the most general supersymmetric background and that all subsequent steps can be reduced to a completely linear system of equations. 

We note that one can, of course, simplify things by taking the base to be hyper-K\"ahler and restricting $\beta$ to be independent of $v$.   In particular, this means that $\psi =0$ and that the equation  (\ref{betacond}) is a simple, linear self-duality condition on the base. 

\subsection{The tensor gauge field}

Supersymmetry requires that the self-dual parts  (in six dimensions)  of the field strength, $G$, and spin connection be the same.  This means that the supersymmetries will be constant in this coordinate system and frame:
\begin{equation} 
\partial_\mu \epsilon ~=~ 0  \,. \label{consteps}
\end{equation} 
The remaining components of the field strength can be written in terms of the dilaton and $2$-from, $K$, that is self-dual on the four-dimensional base.   The final expression may be found in  \cite{Cariglia:2004kk}:
\begin{eqnarray}  
e^{\sqrt{2} \phi} \, G  &=& \coeff{1}{2}\, *_4 \big( DH  + H \, \dot \beta  - \sqrt{2} \, H\, D \phi \big) \nonumber  \\
  &&  -\coeff{1}{2}\,  e^+ \wedge e^-  \wedge \big( H^{-1} DH  +  \, \dot \beta  + \sqrt{2} \,  D \phi \big) \nonumber  \\
  && -  e^+ \wedge \big( - H \psi  + \coeff{1}{2}\,  (D\omega)^-  - K    \big)   + \coeff{1}{2}\, H^{-1}  \, e^- \wedge D\beta \,,  \label{Gform}   
\end{eqnarray}
and, for completeness, we give:
\begin{eqnarray}  
e^{\sqrt{2} \phi} \,*_6 G  &=& \coeff{1}{2}\, *_4 \big( DH  + H \, \dot \beta  +  \sqrt{2} \, H\, D \phi \big)  \nonumber  \\
  &&  -\coeff{1}{2}\,  e^+ \wedge e^-  \wedge \big( H^{-1} DH  +  \, \dot \beta  -  \sqrt{2} \,  D \phi \big)  \nonumber  \\
  && -  e^+ \wedge \big(-  H \psi  + \coeff{1}{2}\,  (D\omega)^- +  K    \big)    + \coeff{1}{2}\, H^{-1}  \, e^- \wedge D\beta \,. \label{dualGform}   
\end{eqnarray}
where we define:
\begin{equation}
(D \omega)^\pm ~\equiv~  \coeff{1}{2}\, (D \omega \pm *_4 D \omega) \label{ompmdefn}   \,.
\end{equation}
Our expressions for (\ref{Gform})  and (\ref{dualGform}) differ from those of \cite{Cariglia:2004kk}  by replacing  $\psi \to - \psi$  as a result of the difference in conventions outlined above.

The Bianchi identity  and the equation of motion are simply:
\begin{equation}
d\, G~=~  0\,, \qquad d \big(e^{2 \sqrt{2} \phi}\, *_6 G \big) ~=~  0  \label{Geqna}   \,,
\end{equation}
and they imply the following differential identities\footnote{We have also corrected sign errors in equation (4.19) of  \cite{Cariglia:2004kk} that would otherwise render it inconsistent with \cite{Gutowski:2003rg} once one has replaced $\psi \to - \psi$.}:
\begin{align}  
D\big(H^{-1}  e^{\sqrt{2} \phi} (K - H\cG - H \psi) \big)   &~+~ \coeff{1}{2} \,\partial_v *_4 \big(D( H\,  e^{\sqrt{2} \phi}) +  H\,  e^{\sqrt{2} \phi}  \dot \beta \big)   \nonumber \\
& ~-~  H^{-1}  e^{\sqrt{2} \phi} \, \dot \beta\wedge (K - H\cG -H  \psi) ~=~0 \,, \label{Bianchi1}   \\
-D\big(H^{-1}  e^{-\sqrt{2} \phi} (K + H\cG +H \psi) \big)   &~+~ \coeff{1}{2}\, \partial_v *_4 \big(D( H\,  e^{-\sqrt{2} \phi}) +  H\,  e^{-\sqrt{2} \phi}  \dot \beta \big)   \nonumber \\
& ~+~  H^{-1}  e^{-\sqrt{2} \phi} \, \dot \beta\wedge (K+ H\cG + H \psi) ~=~0 \,,\label{Bianchi2}  
\end{align}
and
\begin{eqnarray}  
D *_4 \big[D( H\,  e^{\sqrt{2} \phi}) +  H\,  e^{\sqrt{2} \phi}  \dot \beta\, \big]  &=& 2\, H^{-1} e^{\sqrt{2} \phi}  (K -   H\cG) \wedge D\beta \,,\label{Geom1}   \\
D *_4 \big[D( H\,  e^{-\sqrt{2} \phi}) +  H\,  e^{-\sqrt{2} \phi}  \dot \beta \, \big]  &=&  -2\, H^{-1} e^{-\sqrt{2} \phi} (K  + H\cG) \wedge D\beta  \,. \label{Geom2}  
\end{eqnarray}
where $\cG$ is defined as in  \cite{Cariglia:2004kk}\footnote{This differs by a factor of two compared to the definition in  \cite{Gutowski:2003rg}.}:
\begin{equation}
\cG  ~\equiv~  \frac{1}{2 H} \big[\, (D \omega)^+ ~+~  \coeff{1}{2}\, \cF \, D\beta\, \big] \label{cGdefn} \,.
\end{equation}
Note that this form is self-dual in four dimensions.

\subsection{The linear structure}

To expose the linear structure, one first defines the following forms and functions
\begin{equation}
\Theta_1 ~\equiv~  H^{-1} e^{-\sqrt{2}\phi} (K+H \cG + H \psi)\,, \qquad \Theta_2 ~\equiv~ H^{-1} e^{\sqrt{2}\phi} (-K+H \cG + H \psi)\,, \label{ThetaDefns}
\end{equation}
\begin{equation}
Z_1 ~\equiv~H\, e^{\sqrt{2}\phi}\,,\qquad Z_2 ~\equiv~ H\, e^{-\sqrt{2}\phi}\,.  \label{ZDefns}
\end{equation}
Note that  the $\Theta_j$, $j=1,2$, are almost self-dual:
\begin{eqnarray}
*_4 \Theta_1   &=& \Theta_1  -   2 \,  e^{-\sqrt{2}\phi}\, \psi ~=~ \Theta_1  -  2 \,  Z_2\, \hat \psi  \,,  \label{Theta1dual} \\
*_4 \Theta_2   &=& \Theta_2  -   2 \,  e^{\sqrt{2}\phi}\, \psi ~=~ \Theta_2  -  2 \,  Z_1\, \hat \psi  \,.  \label{Theta2dual}
\end{eqnarray}
In particular, the anti-self-dual parts of the $\Theta_j$ are proportional to $\hat \psi$.

With these definitions, equations (\ref{Bianchi1}) and (\ref{Geom1}) become
\begin{equation}
\tilde d \Theta_2 ~=~  \partial_v  \big[ \coeff{1}{2} *_4 (D Z_1 + \dot{\beta} Z_1)~+~  \beta \wedge \Theta_2 \big] \label{Theta2eqna}\,,
\end{equation}
\begin{equation}
D *_4 (D Z_1 + \dot{\beta} Z_1) = - 2 \,\Theta_2 \wedge  D \beta\,, \label{Z1eqn}
\end{equation}
while (\ref{Bianchi2}) and (\ref{Geom2}) become
\begin{equation}
\tilde d \Theta_1~=~  \partial_v  \big[ \coeff{1}{2} *_4 (D Z_2 + \dot{\beta} Z_2)~+~  \beta\wedge \Theta_1 \big] \label{Theta1eqna} \,,
\end{equation}
\begin{equation}
D *_4 (D Z_2 + \dot{\beta} Z_2) = - 2 \,\Theta_1 \wedge  D \beta\,. \label{Z2eqn}
\end{equation}
Note that these equations and the duality properties of the $\Theta_j$ also imply:
\begin{equation}
\tilde d (*_4 \Theta_1)~=~  -2\, \tilde d (Z_2 \hat \psi) +  \partial_v  \big[ \coeff{1}{2} *_4 (D Z_2 + \dot{\beta} Z_2)~+~  \beta\wedge \Theta_1 \big] \label{Theta1eqnb} \,,
\end{equation}
and
\begin{equation}
\tilde  d (*_4 \Theta_2)~=~  -2\, \tilde d (Z_1 \hat \psi)  + \partial_v  \big[ \coeff{1}{2} *_4 (D Z_1 + \dot{\beta} Z_1)~+~  \beta \wedge \Theta_2 \big] \label{Theta2eqnb}\,,
\end{equation}
Since $\beta$ and $\hat \psi$ are known fields on the base manifold, equations (\ref{Theta2eqna}),  (\ref{Z1eqn}) and  (\ref{Theta2eqnb}) define a \emph{linear system} for $\Theta_2$ and $Z_1$, while equations (\ref{Theta1eqna}),  (\ref{Z2eqn}) and  (\ref{Theta1eqnb}) define a \emph{linear system} for $\Theta_1$ and $Z_2$.

It follows that once we have fixed the base manifold, $\cB$, and the vector field $\beta$, the $Z_j$ and $\Theta_j$ are   determined by linear equations.  One can then invert  (\ref{ZDefns}) and (\ref{ThetaDefns}) to determine $H$, $\phi$, $K$ and $\cG$.

\subsection{The angular momentum vector and the  last metric function}

The last two equations that determine the supersymmetric background come from inverting  (\ref{cGdefn}):
\begin{equation}
 (D \omega)^+  ~=~   2\, H\, \cG ~-~   \coeff{1}{2} \cF \, D\beta   \label{cGdefninv} \,.
\end{equation}
and from the one Einstein equation that is not implied by the supersymmetry conditions.

To write the latter equation it is convenient to define:
\begin{equation}
L ~\equiv~  \dot{\omega}~+~ \coeff{1}{2}\, \cF \dot \beta ~-~  \coeff{1}{2}\,  D \cF\,.   \label{Ldefn}
\end{equation}
Note that this is gauge invariant under the transformation:
\begin{equation}
 \cF ~\to~  \cF +2\, \partial_v f   \,, \qquad  \omega ~\to~ \omega + D f  \,,   \label{gaugetrf}
\end{equation}
for some function, $f(v, x^m)$.  This transformation is induced by a
coordinate change $u \to u + f(v, x^m)$ in the metric (\ref{sixmet}).
The last necessary equation may then be written \cite{Cariglia:2004kk}:
\begin{eqnarray}
*_4 D *_4 L &=& \coeff{1}{2}\, H h^{mn}\partial_v^2 (H h_{mn}) + \coeff{1}{4}\, \partial_v (H h^{mn}) \,\partial_v (H h_{mn}) - 2\, \dot{\beta}_m\, L^m + 2\, H^2 \,\dot{\phi}^2  \nonumber\\
&& - \coeff{1}{2} \, H^{-2} \big(D\omega +  \coeff{1}{2} \,  \cF D\beta\big)^2  + 2\, H^{-2} \big(K - H\, \psi + \coeff{1}{2} (D\omega)^- \big)^2 \,. \label{lasteqn}
\end{eqnarray}
where, for any $2$-form, $\cM^2 = \frac{1}{2} M_{mn} M^{mn}$.

One can now rewrite this apparently  non-linear equation in $\omega$ by first replacing $D \omega$ by $(D\omega)^- +  (D\omega)^+$ and expanding the squares.  The $((D\omega)^-)^2$ terms cancel and one can then replace $(D\omega)^+$ using (\ref{cGdefninv}) and the definitions of the $\Theta_j$'s.  Further simplifications can then be made by using the duality properties of all $2$-forms and the fact that the contraction of a self-dual and anti-self dual form vanishes identically.   We then find that (\ref{lasteqn}) can be rewritten as:
\begin{eqnarray}
*_4 D *_4 L &=& \coeff{1}{2}\, H h^{mn}\partial_v^2 (H h_{mn}) + \coeff{1}{4}\, \partial_v (H h^{mn}) \,\partial_v (H h_{mn}) - 2\, \dot{\beta}_m\, L^m + 2\, H^2 \,\dot{\phi}^2  \nonumber\\
&& - 2*_4 \Big[\, \Theta_1 \wedge \Theta_2 ~-~  H^{-1}  \psi  \wedge D \omega \, \Big] \,.  \label{Feqn}
\end{eqnarray}
In re-writing the right-hand side of (\ref{Feqn}) in terms of the dual of  wedge products of two forms, rather than the contractions of two forms, it is important to remember that the latter  involves the {\it sum} of the  contractions of the self-dual and anti-self-dual parts, whereas  former produces the {\it difference} between the contractions of the self-dual and the anti-self-dual parts.  Thus one needs to employ the duality properties (\ref{Theta1dual}) and  (\ref{Theta2dual}) and the the anti-self-duality of $\psi$ to arrive at the correct expression for   (\ref{Feqn}).

Finally, (\ref{cGdefninv}) can be rewritten as: 
\begin{eqnarray}
D \omega + *_4 D \omega  &=&    2\,Z_1\, \Theta_1 +  2\,Z_2\, \Theta_2  ~-~  \cF \, D\beta - 4\, H\, \psi \nonumber \\
&=&    2\,Z_1\, \big (\Theta_1 - Z_2\, \hat \psi\big) +  2\,Z_2\, \big(\Theta_2 - Z_1\, \hat \psi\big) ~-~  \cF \, D\beta      \label{angmomeqn} \,.
\end{eqnarray}
In the second identity we have arranged the right hand side into manifestly self-dual combinations that follow from  (\ref{Theta1dual}) and   (\ref{Theta2dual}).

Note that   (\ref{Feqn}) and (\ref{angmomeqn}) are \emph{linear} in $\cF$ and $\omega$. Since  $h_{mn}, H, \phi$, $ \Theta_1,\Theta_2$, $\beta$ and $\psi$ are determined by the base and by solving the earlier linear systems, we see that  (\ref{Feqn}) and (\ref{angmomeqn})  define another \emph{linear system} that can be used to determine $\cF$ and $\omega$.
Thus, once one has fixed the base geometry and one has determined $\beta$, the entire solution is determined by a linear structure.  In particular, vast numbers of new solutions can be obtained by superposition.

\section{Simplifying  the fluxes and flux equations}
\label{sect:susyeqns}

Define the $3$-forms
\begin{equation}
\gamma_1 ~\equiv~   \coeff{1}{2} *_4 (D Z_1 + \dot{\beta} Z_1)~+~  \beta \wedge \Theta_2 \,, \qquad
 \gamma_2 ~\equiv~   \coeff{1}{2} *_4 (D Z_2 + \dot{\beta} Z_2)~+~  \beta \wedge \Theta_1 \,.  \label{gammadefn} 
\end{equation}
Then  equations (\ref{Theta2eqna}),   (\ref{Z1eqn}), (\ref{Theta1eqna}) and  (\ref{Z2eqn}) may be written very simply as:
\begin{equation}
\tilde d \Theta_2 ~=~  \partial_v  \gamma_1  \,, \qquad \tilde d \gamma_1 ~=~ 0 \,;   \qquad \tilde d \Theta_1 ~=~  \partial_v  \gamma_2  \,, \qquad \tilde d \gamma_2 ~=~ 0 \,.  \label{ZThetaeqns}
\end{equation}

The tensor gauge field (\ref{Gform}) and its dual  (\ref{dualGform}) can be written in terms of electric and magnetic parts:
\begin{eqnarray}  
G  &=&  d \big[ - \coeff{1}{2}\,Z_1^{-1}\,(du + \omega) \wedge (dv + \beta)\, \big] ~+~ \widehat G_1 \,,  \label{niceGform}   \\
e^{2\, \sqrt{2} \phi} \,*_6 G  &=&   d \big[ - \coeff{1}{2}\,Z_2^{-1}\,(du + \omega) \wedge (dv + \beta)\,   \big] ~+~ \widehat G_2  \,,\label{nicedualGform}   
\end{eqnarray}
where 
\begin{eqnarray}  
 \widehat G_1  &\equiv&   \coeff{1}{2} *_4 (D Z_2 + \dot{\beta} Z_2) ~+~  (dv+ \beta) \wedge \Theta_1 \,,  \label{G1hat}   \\
 \widehat G_2 &\equiv&   \coeff{1}{2} *_4 (D Z_1 + \dot{\beta} Z_1) ~+~  (dv+ \beta) \wedge \Theta_2\,. \label{G2hat}   
\end{eqnarray}
In particular, the Bianchi identities and Maxwell equations then require the closure of the $\widehat G_j$, which means these quantities do indeed measure the conserved magnetic charge.  

Computing $d  \widehat G_j$ one easily finds:
\begin{eqnarray}
 D\big[ *_4 (D Z_i + \dot{\beta} Z_i) \big] ~+~ 2\,(D\beta)\wedge \Theta_j &=& 0 \,,  \label{eoms1}\\ 
 D \Theta_j -\dot \beta \wedge  \Theta_j ~-~   \partial_v \big[\coeff{1}{2} *_4 (D Z_i + \dot{\beta} Z_i) \big] &=&   0  \,.  \label{eoms2}
\end{eqnarray}
where $\{i,j\} =\{1,2\}$.  One can easily check that these equations are equivalent to (\ref{ZThetaeqns}).

The closure of $\widehat G_j$ also means that they are locally exact:
$\widehat G_j = d \xi_j$.  Moreover, because all the fields are
independent of $u$ and there are no $du$ terms in $\widehat G_j$, one
can assume that the $2$-forms, $\xi_j$, also do not involve any $du$
terms.  One can then make a gauge choice for the $\xi_j$ so as to remove
all the $dv$ terms\footnote{Strictly speaking one can only do this for
the non-trivial Fourier modes along the $v$-direction.  The computation
for the zero-mode will, of course, reproduce the usual five-dimensional
BPS equations.}.  This means that one can write:
\begin{equation}
\coeff{1}{2} *_4 (D Z_i + \dot{\beta} Z_i) ~=~  D \xi_j  \,, \qquad   \Theta_j ~=~ \partial_v\, \xi_j    \,.  \label{exact1}
\end{equation}
The integrability conditions for $\xi_j$ are, of course, the equations (\ref{eoms1}) and (\ref{eoms2}).

Now recall the duality properties, (\ref{Theta1dual}) and (\ref{Theta2dual}), of the $\Theta_j$.  These imply that the $\Theta_j$ can be written entirely in terms of three arbitrary functions on the base and if one expands in Fourier modes, one can use the second equation in (\ref{exact1}) to write the $\xi_j$ in terms of three arbitrary functions.  The first equation in (\ref{exact1}) is, in fact, four independent equations, one for each component, and so these equations can, in principle, be used to determine the four unknown functions: $Z_i$ and the three functions in $\xi_j$.

\section{Examples}
\label{sect:examples}

In this section, we use our linear procedure to construct several new
solutions.  These solutions can be thought of as configurations of D1
and D5 branes in type IIB string theory compactified to six dimensions.
Indeed, the unconstrained two-index gauge field $B_{\mu\nu}$ of the
six-dimensional theory corresponds to the RR 2-form, $C_{\mu\nu}$, and
this is sourced by D1-branes and D5-branes.  The scalar, $\phi$,
corresponds to $\Phi\over\sqrt{2}$ where $\Phi$ is the ten-dimensional
dilaton.

\subsection{The D1-P supertube spiral}

The first solution corresponds to the bound state of D1 branes along the
$v$ direction and momentum (P) along the $v$ direction.
This D1-P system is dual to the well-known F1-P system
\cite{Dabholkar:1989jt} and the bound state is described by the D1
world-volume moving along $\vec{x}=\vec{F}(v)$ where $\vec{x}=(x^1,\dots,x^4)$ and
$\vec{F}(v)$ is an arbitrary vector function.
This solution is not new, but casting it in our formalism is instructive for the construction of the more general solutions described in the next subsection.

The metric and RR 2-form can be obtained by dualizing the
known supergravity solution of the F1-P system \cite{Dabholkar:1995nc,
Lunin:2001fv} and putting it into our form.  The resulting fields are:
\begin{align}
\label{solD1P}
\begin{split}
 h_{mn}&=\delta_{mn},\qquad
 \beta=\psi=0,\\
 Z_1&=1+H_1,\qquad H_1={Q_1\over |\vec x-\vec F(v)|^2},\qquad Z_2=1,\\
 \Theta_1&=0,\qquad
 \Theta_2 ={1\over 2}(1+*_4)\,\tilde{d}(H_1 \dot{F}_m dx^m),\\
 \omega&
 =H_1\dot{F}_m  dx^m,\qquad
 \cF =-H_1 \dot{F}^2,\qquad
 K=-{1\over 2}\Theta_2.
\end{split}
\end{align}
The base space of this solution is flat $\IR^4$.  The one-form, $\beta$,
is sourced by KK monopole charges whose special direction is $v$ and,
since the solution has no such charge, $\beta$ vanishes. Hence, the $v$ fiber is trivial and the spatial metric is a warped of $\IR^4 \times S^1$.  Because the D1
world-volume is not straight but along the curve $\vec{x}=\vec{F}(v)$,
there are D1 charges along $v$ as well as $\vec{x}$, and they are
respectively encoded in $Z_1$ and $\Theta_2$.  Also, because the D1
world-volume is moving, there are momentum charges along $v$ and $\vec{x}$,
and they are respectively encoded in $\cF$ and $\omega$.

The foregoing solution involves only one ``strand'' of D1-branes, but one can easily include multiple strands by summing over strands:
\begin{align}
 Z_1=1+\sum_{p=1}^n {Q_{1p}\over |\vec x-\vec F_p(v)|^2}.\label{multiplestrandsD1P}
\end{align}
Each function, $\vec F_p(v)$,  is arbitrary and describes  the position of the $p^{\rm th}$ strand.  The other fields, $\Theta_2$, $\omega$ and $\cF$, are given by expressions similar to \eqref{solD1P} but now including summation over multiple strands.

\subsection{D1-D5-P supertube}
\label{ss:D1D5P->d1d5:single strand}

It is relatively simple to add D5-branes extending along the $v$ direction\footnote{The D5-branes also wrap  the
four internal directions of the compactification of the IIB theory to six dimensions.} and obtain a more general traveling-wave solution.  
This solution will have D1, D5 and P charges, and describes a traveling momentum wave on a D1-D5 system. The resulting arbitrarily-shaped profile is a three-charge two-dipole charge supertube spiral, and our solution describes an arbitrary distribution of such parallel spirals. To our knowledge this solution is new.
As discussed in \cite{Bena:2011uw},  this configuration corresponds to the first of the two supertube transitions involved in the  ``double bubbling:''   the D1-D5-P system is polarized into a three-charge supertube spiral of arbitrary shape. 

Just as in the D1-P example, we take the base to be flat and, because this tube will have no KKM dipole moment, we again set $\beta$ to zero\footnote{Adding the KKM dipole moment along a second, independent profile represents the second supertube transition of  the``double bubbling'' and would greatly increase the technical challenge of finding the solution.}. Hence 
\begin{align}
  h_{mn}=\delta_{mn},\qquad \beta=\psi=0\label{betapsi0}\, ,
\end{align}
which implies that the $Z_i$ equations and $\Theta_i$ equations decouple.  The equations \eqref{Z1eqn}, \eqref{Z2eqn} simply give the Laplace equation on the base:
\begin{align}
 \tilde d *_4 \tilde d \, Z_i =0,\qquad i=1,2.
\end{align}
Following the D1-P example, we choose the following harmonic functions:
\begin{align}
 Z_i=c_i+\sum_{p=1}^n{Q_{ip}\over |\vec x-\vec F(v)-\vec a_p|^2}
\label{ZansatzD1D5P}
\end{align}
with constants, $c_i$, an arbitrary vector function, $\vec{F}(v)$, and
\emph{constant} vectors, $\vec{a}_p$.  Because singularities of $Z_1$
and $Z_2$ represent the D1 and D5 charges, respectively, this solution
corresponds to $n$ \emph{parallel} strands of D1-D5 world-volume along
$\vec{x}=\vec{F}(v)+\vec{a}_p$.  For later convenience, we define:
\begin{align}
 H_i=\sum_{p=1}^n {Q_{ip}\over |\vec x-\vec F(v)-\vec a_p|^2} \,, \qquad i =1,2 \,.
\end{align}

The magnetic components, $\Theta_i$,  of the solution are determined by \eqref{Theta2eqna} and \eqref{Theta1eqna}, which now become:
\begin{align}
 \tilde d \Theta_i={1\over 2}*_4\tilde{d} \dot{Z}_j,
\qquad \{i,j\}=\{1,2\}.
 \label{xdj5Aug11}
\end{align}
Their solution is simply:
\begin{align}
 \Theta_i={1\over 2}(1+*_4)\,\tilde d (H_j \dot{F}_m dx^m) \, .
 \label{Thetaex}
\end{align}
One also has the freedom to add  closed forms  to $\Theta_i$.  The functions $H,\phi,K$ are
given by
\begin{align}
 H=Z_1^{1/2}Z_2^{1/2} ,\qquad   e^{2\sqrt{2}\phi}={Z_1\over Z_2},\qquad  K={1\over 2}(Z_1\Theta_1-Z_2\Theta_2),
\end{align}

Since the $ \Theta_i$ are not closed, they do not define conserved
magnetic charges. Indeed, as noted in Section \ref{sect:susyeqns}, the
conserved magnetic charges of six-dimensional solutions are defined by
integrating the $3$-forms, $\widehat G_i$, defined in
(\ref{G1hat}) and (\ref{G2hat}), over three cycles.  The only
non-trivial $3$-cycles in our D1-D5 strands are the Gaussian $S^3$'s
that surround each strand in the five-dimensional spatial metric.  The
corresponding flux integrals then give the only quantized charges in the
solution: the $Q_{jp}$.  On the other hand, one can imagine smearing the
solution along the $v$ fiber and compactifying along that direction.
One now has a profile along the base and a Gaussian $S^2$ surrounding
that profile.  This means that there is a non-trivial $3$-cycle defined
by this $S^2$ and the $v$ fiber.  The integral of $\widehat G_i$ over
this cycle reduces to the integral of $ \Theta_i$ over the $S^2$ and,
from (\ref{Thetaex}), the magnetic charges of the smeared profile
defined in this way will be proportional to $Q_{ip} |\dot { F}|$.  Thus
the smeared versions of our new solutions\footnote{While we are not
explicitly smearing the solution here, the second supertube transition
in double bubbling does smear the solution and these considerations will
become important in future work.}  will have electric charges, $Q_{ip}$,
and magnetic dipole charges given by $Q_{ip} |\dot { F} |$.

With the simple background defined by \eqref{betapsi0}, the $\omega$ and
$L$ equations are decoupled and the $\omega$ equation \eqref{angmomeqn}
becomes
\begin{align}
 (1+*_4)\tilde d\omega ~=~ 2(Z_1\Theta_1+Z_2\Theta_2).\label{xgb5Aug11}
\end{align}
It is easy to check that the following is a solution:
\begin{align}
 \label{omega_D1D5P}
  \omega
 ~=~  (H_1H_2+c_1 H_2+c_2 H_1)\dot{F}_m dx^m.
\end{align}
This has a pole structure familiar from the 3-charge 2-dipole-charge
black rings, but now everything depends on $v$ as well.  Again, in
\eqref{omega_D1D5P}, we could have added a closed form to
$\omega$. 

Finally, $\cF$ is determined by \eqref{Feqn}, which can be written as
\begin{align}
 *_4\tilde d *_4 \tilde d \,\cF=
 -\partial_m \partial_m \cF
 &=
 2*_4\tilde d *_4 \dot \omega 
 -2 (\dot Z_1 \dot Z_2+Z_1\ddot Z_2+\ddot Z_1 Z_2)
 +4*_4(\Theta_1\wedge \Theta_2)\notag\\
 &=2\, \partial_m H_1 \, \partial_m H_2\,  \dot{F}^2.
\end{align}
In deriving this expression and some of the expressions above, the
relation $\dot{Z}_i=\dot{H}_i=-\dot{F}_m\,\partial_m
Z_i=-\dot{F}_m\,\partial_m H_i$ is useful.  Because $H_i$ are harmonic
with respect to $\vec{x}$, we find
\begin{align}
 \cF&~=~-H_1H_2\,\dot{F}^2 ~+~ H_3 \,,\label{cF_D1D5P}
\end{align}
where we have chosen to add explicitly a function, $H_3(v,\vec{x})$,
that is harmonic on the four-dimensional base.  This function
corresponds to putting a freely-choosable momentum profile on the D1-D5
world-volume.

When we only have one strand ($n=1$), with equal D1 and D5 charges and harmonic functions ($c_1=c_2=1$, $Q_1=Q_2$), then our solution reduces
to the black string with traveling waves of \cite{Horowitz:1996th,Horowitz:1996cj}. The closed form that we could have added in \eqref{omega_D1D5P} corresponds to the
angular momentum distribution discussed in \cite{Horowitz:1996cj} while
the harmonic form $H_3$ in \eqref{cF_D1D5P} corresponds to the momentum
distribution studied in \cite{Horowitz:1996th}. Our general solution describes a parallel distribution of black strings whose traveling waves have identical profiles. For a generic momentum distribution the strings will have a nonzero horizon area, and when the source of $H_3$ is such that the classical horizon area is zero, the black strings with traveling waves become three-charge two-dipole-charge supertube strands. 

It is interesting to note that the solution for two-charge supertube spirals presented in the previous subsection and the solution for multiple D1-P and D5-P strands and black strings with traveling waves given here both have a certain ``action at a distance'' property:  if one slices the solution at a constant value of $v$, then the fields on that  slice only depend on the positions of the strands and on their $v$-derivatives on that slice, but are completely independent of the behavior of the strands at any other values of $v$!   Hence, if one thinks of the supertube strands as traveling waves on D1 and D5 branes, the fields at some location ($x_i, v_0$) far away from the strand are independent of the values of the functions at any point except $v_0$.

This ``action at a distance'' behavior gives some insight into why the solution with parallel strands is much more straightforward to find than solutions with different profiles for each strand.  If one takes constant-$v$ slices of a solution in which all the profiles are the same then, in any slice, the distance between the strands and the derivatives of the $F_i$ are the same.  Since these derivatives give the dipole charges, it is not hard to see that this implies that, in any particular slice, the symplectic product between the charges of any two centers is zero.  Hence, there are no bubble equations, or integrability conditions, to govern the separation between parallel strands and these strands can be moved at any position, as long as they remain parallel. 

To illustrate this, it is easier to consider a D1-P strand and a parallel  D5-P strand. In a given slice, the first strand has a D1 dipole charge of magnitude and orientation:
\begin{equation}
d_1 ~=~  Q_1  \dot{\vec F} (v)
\end{equation}
and the second strand has a D5 dipole charge:
\begin{equation}
d_5 ~=~  Q_5 \dot{\vec F} (v) \,.
\end{equation}
Hence, the symplectic product between the two strands, $Q_1 d_5 - Q_5 d_1$, vanishes exactly in any slice. Given that D1-D5-P three-charge spirals can always be decomposed into parallel D1-P and D5-P strands \cite{Bena:2011uw}, and given that D1-P strands do not interact with each other (as explained in the previous section) the foregoing calculation indicates that parallel strands can be moved anywhere in the solution and there are no constraints on their position. One can make a similar argument for parallel black strings with identical traveling waves: they only differ from the three-charge supertube strands in the extra ``momentum'' harmonic function $H_3$, and since the solution has no KKM charges this extra harmonic function does not affect the bubble equations. The absence of bubble equations can also be understood from the absence of Dirac-Misner strings in the expression,   \eqref{omega_D1D5P}, for $\omega$.  

If we do not assume that different strands are parallel then the linear procedure still gives  straightforward solutions for the electric potentials, or warp factors: one simply replaces $\vec F(v)$ in  \eqref{ZansatzD1D5P} by $\vec F_p(v)$.  However,  the solution for $\omega$ and $\cF$ will be significantly more complicated.   Moreover, the dipole charges at various  points on each strand will not be parallel to each other, and there will be non-trivial $v$-dependent bubble equations.   That is,  requiring the absence of Dirac strings and the  absence of closed time-like curves will  impose conditions on the locations of the strands on every $v$-slice.    Based on a rough counting, we expect the solution for two strands to be given by seven arbitrary functions: four shapes modes for each strand minus one for the $v$-dependent bubble equation.

\section{An Intermediate Class of Solutions}
\label{sect:simpler}

We have found that the system of equations describing supersymmetric
solutions of six-dimensional, minimal, ungauged supergravity coupled to
an anti-symmetric tensor multiplet can be solved in a linear process
once the base metric and the vector field, $\beta$, have been
determined. Since the vector field, $\beta$, is now a \emph{geometric}
field, it may be viewed as defining the background spatial geometry and
so the only non-linearities lie in determining this background geometry
while the remaining parts of the solution are entirely linear.

  When the solutions to this theory do not depend on the ``common D1-D5'' direction, $v$, they can be reduced to solutions of yields $\cN=2$ ungauged five-dimensional supergravity coupled to two vector multiplets, and the complicated linear system found here collapses to the simpler linear system found in \cite{Bena:2004de}. 
 
There is, however,  a very interesting intermediate class of solutions in which  $\beta$, the complex structures and the base metric are $v$-independent.  The condition (\ref{Jcond}) then requires the complex structures to be closed and thus  the base metric must, in fact,  be hyper-K\"ahler.    One also has  $\hat \psi \equiv 0$.   The system of equations then collapses to:  
\begin{equation} 
\tilde d \beta ~=~  *_4 \tilde d \beta\,, \qquad *_4 \Theta_1   ~=~ \Theta_1\,, \qquad *_4 \Theta_2  ~=~ \Theta_2     \,, \label{sdforms}
\end{equation} 
\begin{equation}
\tilde d \Theta_2 ~-~  \beta\wedge \dot \Theta_2  ~=~   \coeff{1}{2} *_4  D \dot Z_1  \,, \qquad D *_4  D Z_1  ~=~  - 2 \,\Theta_2 \wedge  \tilde d  \beta\,,  \label{Z1Theta2eqns} 
\end{equation}
\begin{equation}
\tilde d \Theta_1  ~-~  \beta\wedge \dot \Theta_1   ~=~   \coeff{1}{2} *_4  D  \dot Z_2   \,, \qquad D *_4  D Z_2   ~=~  - 2 \,\Theta_1 \wedge   \tilde d   \beta  \,,  \label{Z2Theta1eqns} 
\end{equation}
\begin{equation}
*_4 D *_4 (\dot{\omega}  -  \coeff{1}{2}\,  D \cF)   ~=~        
 - 2*_4   \Theta_1 \wedge \Theta_2  ~+~ Z_1 \partial_v^2 Z_2 ~+~ Z_2 \partial_v^2 Z_1 +  (\partial_v Z_1)(\partial_v Z_2) \,,  \label{Feqnsimp}
\end{equation}
\begin{equation}
D \omega + *_4 D \omega  ~=~   2\,Z_1\, \Theta_1 +  2\,Z_2\, \Theta_2  ~-~  \cF \, \tilde d \beta      \label{angmomeqnsimp} \,.
\end{equation}
This linear system is similar to that of \cite{Bena:2004de}, except that all the functions and fields, with the exception of $\beta$, are now allowed to be functions of the fiber coordinate, $v$.  We believe that even with the simplifying assumptions above there are going to be rich new varieties of microstate geometries in six dimensions.

\section{Conclusions}
\label{sect:conclusions}

Six-dimensional, minimal ungauged supergravity coupled to an
anti-symmetric tensor multiplet, when reduced to five dimensions, yields
$\cN=2$ supergravity coupled to two vector multiplets.  Thus the
six-dimensional theory is extremely important to the general study of
microstate geometries: smooth geometries that have the same
asymptotic structure at infinity as a given five-dimensional black hole
or black ring.  Generic supersymmetric solutions in six-dimensions 
are expected to have a much richer structure than their five-dimensional
counterparts, not only because such solutions can depend upon the
compactification directions but also because we expect to find genuinely
new, smooth supersymmetric solitons, like the superstratum
\cite{Bena:2011uw}.

We have shown that the system of equations underlying all supersymmetric solutions in this six-dimensional theory reduces to a linear system of equations after one has made the first step of laying down some of the geometric elements of the metric.  Specifically, the six-dimensional metric must have the form  (\ref{sixmet}) and the four-dimensional base is required to be almost hyper-K\"ahler with $2$-forms  satisfying (\ref{Jalg}).   On this base manifold one must choose a vector field, $\beta$, whose field strength is self-dual, (\ref{betacond}), and that satisfies the differential constraint (\ref{Jcond}).  This vector field is also part of the five-dimensional spatial geometry because it determines how the $S^1$ is fibered over the spatial base.  The choices of the base metric and  the vector field, $\beta$, represent the only non-linearities in the system of equations.   After this, all of the electric potentials, warp factors, magnetic fluxes and the angular momentum vector are determined by linear systems of equations.  Moreover, if $\beta$ is taken to be independent of the fiber coordinate, $v$, then it is simply a self-dual $U(1)$ gauge field and its equation of motion is also linear.  

To summarize the process in more detail, once one has chosen an almost hyper-K\"ahler base metric on the four manifold and has fixed $\beta$, one uses the background to construct $\hat \psi$ as defined in  (\ref{psidefn}).  One then  solves the coupled linear systems  (\ref{Theta2dual}),   (\ref{Theta2eqna})  and   (\ref{Z1eqn}) for $Z_1$ and $\Theta_2$ and   (\ref{Theta1dual}),   (\ref{Theta1eqna})  and  (\ref{Z2eqn}) for $Z_2$ and $\Theta_1$.    The functions $H$ and $\phi$ are then obtained from  (\ref{ZDefns}) and the forms $K$ and $\cG$ are obtained from  (\ref{ThetaDefns}).   With these known sources, $\beta$ and the base metric,   equations (\ref{Feqn}) and (\ref{angmomeqn}) are \emph{linear} in $\cF$ and $\omega$.  As with all linear systems there are choices of homogeneous solutions and these correspond to sources for the fluxes or must be chosen to make the metric suitably regular and remove closed time-like curves.  

It would, of course, be very nice to have some greater control over the general class of solutions, and have a more systematic understanding of the starting point of this linear system: the almost hyper-K\"ahler metrics with self-dual Maxwell fields defined by a vector field, $\beta$, satisfying (\ref{Jcond}).  The self-duality equation for $D\beta$ is non-linear in $\beta$ because of the definition (\ref{betaFS}), however if one expands in Fourier modes one can view this as Euclidean self-dual Yang-Mills where the structure constants are those of  the (classical) Virasoro algebra.  It may therefore be possible to solve this in a straightforward manner using some generalization of the ADHM construction.  Alternatively, there might be an interesting classification of almost hyper-K\"ahler metrics and vector fields $\beta$ if one  assumes a $U(1)$  isometry on the base and that the almost hyper-K\"ahler forms are invariant under this action.  

From past experience we expect our key observation of linearity to make
the future construction of six-dimensional BPS solutions far simpler and
to enable the construction of whole new classes of solutions through the
superposition of others. In addition, the linear structure will also
make the analysis of the moduli space of solutions a much more
straightforward process and enable a much simpler analysis of BPS
fluctuations around a given background.  While the analysis of the
generic almost hyper-K\"ahler base and the non-linear equation for
$\beta$ is, as yet, unsolved, it is worth remembering the observation
made in Section \ref{sect:simpler}: There are already going to be large
new classes solutions that start from a general hyper-K\"ahler base and
that are completely determined by a linear system on that base.

Apart from the fact that the classification of six-dimensional BPS
solutions is interesting, our primary motivation for dissecting the
system of BPS equations is to look for new, smooth BPS solutions and
particularly for the superstratum \cite{Bena:2011uw}.  Here we have made
some important steps in this direction.  We have reduced the system of
BPS equations to a far more manageable form and we have constructed a
new class of three-charge solutions that have undergone the first of the
two supertube transitions that will lead to the superstratum.  We leave
this the completion of this process for future work.


\bigskip
\section*{Acknowledgments}

\noindent
We would like to thank Nikolay Bobev and Cl\'ement Ruef for interesting discussions and Ben Niehoff for some corrections to the manuscript.  The work of IB was supported in part by the ANR grant 08-JCJC-0001-0, and by the ERC
Starting Independent Researcher Grant 240210 - String-QCD-BH\@. The work
of MS was supported in part by Grant-in-Aid for Young Scientists
(Start-up) 23840017 from the Japan Society for the Promotion of Science
(JSPS)\@.  The work of NPW was supported in part by DOE grant
DE-FG03-84ER-40168.  NPW is also very grateful to the IPhT, CEA-Saclay
for hospitality while some of this work was done. IB, MS and NPW would
like to thank the Aspen Center for Physics for providing a very
stimulating environment while this work was largely completed and as a
result this work was also supported in part by the Aspen Center for
Physics National Science Foundation Grant No.~1066293.  \smallskip

\bibliographystyle{toine}

\providecommand{\href}[2]{#2}\begingroup\raggedright\endgroup

\end{document}